\documentclass[pra,aps,preprint]{revtex4}
\usepackage{amsmath,amssymb}
\usepackage{graphicx}
\usepackage{mathrsfs}

\newcommand{\labhl}{\mathscr{H}_{\rm\scriptscriptstyle E}}
\newcommand{\hs}{H_{\rm\scriptscriptstyle S}}
\newcommand{\hsl}{H_{\rm\scriptscriptstyle SE}}
\newcommand{\ds}{\mathcal{D}_{\rm\scriptscriptstyle S}}
\newcommand{\dsl}{\mathcal{D}_{\rm\scriptscriptstyle SE}}
\newcommand{\hl}{H_{\rm\scriptscriptstyle E}}
\newcommand{\heff}{H_{\rm\scriptscriptstyle eff}}
\newcommand{\trl}{{\rm Tr}_{\rm\scriptscriptstyle E}}
\newcommand{\rhos}{\rho_{\rm\scriptscriptstyle S}}
\newcommand{\rhol}{\rho_{\rm\scriptscriptstyle E}^{\rm\scriptscriptstyle eq}}
\newcommand{\trho}{\tilde{\rho}}
\newcommand{\ul}{U_{\rm\scriptscriptstyle E}}
\newcommand{\ket}[1]{\vert #1 \rangle}
\newcommand{\bra}[1]{\langle #1 \vert}

\newcommand{\innerp}[2]{\langle #1 \vert #2 \rangle}

\begin{document}
\title{Emergence of the Born rule in strongly-driven dissipative systems}
\author{Nilanjana Chanda}
\author{Rangeet Bhattacharyya}
\email{rangeet@iiserkol.ac.in}
\affiliation{Department of Physical Sciences, Indian Institute of Science Education and Research
Kolkata,\\
Mohanpur -- 741246, WB, India}

\begin{abstract}

To understand the dynamical origin of the measurement in quantum mechanics, several models have been put
forward which have a quantum system coupled to an apparatus. The system and the apparatus evolve in time and
the Born rule for the system to be in various eigenstates of the observable is naturally obtained.  In this
work, we show that the effect of the drive-induced dissipation in such a system can lead to the Born rule,
even if there is no separate apparatus. The applied drive needs to be much stronger than the
system-environment coupling. In this condition, we show that the dynamics of a driven-dissipative system
could be reduced to a Milburn-like form, using a recently-proposed fluctuation-regulated quantum master
equation [A. Chakrabarti and R. Bhattacharyya, Phys. Rev. A 97, 063837 (2018)]. The system evolves
irreversibly under the action of the first-order effect of the drive and the drive-induced dissipation. The
resulting mixed state is identical to that obtained by using the Born rule. 

\end{abstract}

\maketitle

\section{Introduction}

Born rule provides the outcome of a measurement of an observable on a quantum system \cite{born1926}. This
rule is introduced as a postulate in the axiomatic formulation of quantum mechanics \cite{cohen_qm}.  The
non-analytic nature of the measurement prompted the development of several dynamical models which aimed to
show that the Born rule was but a natural outcome of the time evolution of a coupled system and apparatus;
with much ingenuity involved in constructing the apparatus and the coupling of the system to the apparatus
\cite{von,green,cini,gaveau,hepp,nakazato,curie_weiss_model}. 

Among the earliest attempts to have a dynamical model, von Neumann formulated the measurement process as a
coupling between two quantum systems, one is the observed system and the other is the measuring apparatus;
the observer does not directly measure the system, but infers the state of the system by observing that of
the apparatus (referred to as the pointer variable) \cite{von}. The system and the apparatus evolve together
and reach a steady-state where eigenstates of the system and the apparatus are entangled. Each state of the
apparatus uniquely identifies an eigenstate of the system. The state and the apparatus evolve under a
unitary propagator and as such the collapse of the state function is not within the scope of von Neumann's
model \cite{zurek2003}.

Hepp proposed several dynamical models, of which the Coleman-Hepp 1972 is well-known \cite{hepp}.
They proposed that a measurement could be modeled as a fast particle passing through a long row (assumed
infinite) of non-interacting spins and flipping them one after another to induce an observable macroscopic
signature.  This spin-flipping local potential is referred to as the Coleman-Hepp or AgBr Hamiltonian. The
choice of this specific Hamiltonian results in an exactly solvable model. This model is further extended by
Nakazato and Pascazio where in addition to the free Hamiltonian of the particle, they considered the free
Hamiltonian of the spin array as well \cite{nakazato}. 

Among the other unitary approaches the model by Cini is also exactly solvable and is constructed using a
spin-1/2 particle (as the system) interacting with a spin-L particle as the apparatus. The interaction
Hamiltonian is proportional to $\sigma_z L_z$, where $\sigma_z$ and $L_z$ are, respectively, the
$z$-components of the spin-1/2 and spin L \cite{cini}. However, being completely unitary, this model too
does not describe the collapse.

In general, the measurement in quantum mechanics is an irreversible process. Therefore, the post-measurement
state of a quantum system could be described by a mixed state density matrix. We note that the
irreversibility also arises naturally in open quantum systems or driven-dissipative systems. Thus the need
of an environment in modeling the measurement process was felt, and several dynamical models of measurement
were proposed which use quantum master equations or in more general terms, use the notion of the
environment. 

Zurek in early 80s, showed that von Neumann's scheme may be extended using an apparatus coupled to
the environment \cite{zurek1981, zurek1982}. The environment is composed of many particles, i.e., it
is assumed to have many degrees of freedom. After a combined evolution of the apparatus and the
environment under a suitably chosen coupling, one needs to take a trace over the environment. The
resulting state is a mixed state with apparatus states, (pointer variables) having different
classical probabilities as per the Born rule. This decoherence-assisted process of selection of the
pointer states is named as einselection \cite{zurek1981, zurek1982, zurek2003}.

Motivated by the fact that an open quantum system show irreversible dynamics, Green proposed modeling of
apparatus with coupling to the thermal baths \cite{green}. In this model, a two-level system (TLS) is
considered which is brought into interaction with separate detectors. Each detector has two sets of
oscillators at different temperatures. The oscillators are coupled by an interaction with the particle. As
such, the particle's states could be detected by a temperature change. After the interaction, the particle
is in a mixed state.

Gaveau and Schulman, proposed a variant where the system was a spin-1/2 particle and the apparatus was a
one-dimensional Ising spin model \cite{gaveau}. One spin of the apparatus interacts with the spin-1/2
system. The energy parameters of the apparatus model are an external field and the spin-spin coupling
(exchange coupling). The system and apparatus evolve together to reach specific steady-states. By observing
the apparatus, one could infer the system's state.

For a static particle, Allahverdyan and others proposed a similar model named Curie-Weiss model
\cite{curie_weiss_model}. In this model, the system is a spin-1/2 whose $z$-component is measured through
coupling with an apparatus that includes a magnet formed by a set of N ($\gg 1$) spins-1/2 coupled to one
another through quartic infinite-range Ising interaction (magnetization in the $z$-direction acts as the
pointer variable) and a phonon bath that interacts with the magnet through a spin-boson coupling.
The reduced density matrix of the spin-1/2 particle dynamically evolves to the form predicted by the
Born rule, with the magnet (the apparatus) reflecting the state of the particle.

In all the approaches described above, the system and the apparatus are made to evolve together. The
apparatus is modeled often in a rather elaborate way, such as in the Curie-Weiss model. Now, with the
increasing importance of quantum information processing, incorporating such a measuring apparatus in a
quantum circuit becomes cumbersome, since each qubit would require its own apparatus. In this work, we show
that it is possible to have a dynamical model without an explicit invocation of an apparatus, provided one
uses the recently-observed drive-induced dissipation within the framework, as described below.

Driven-dissipative dynamics with non-Bloch behavior, which is a manifestation of the drive-induced
dissipation, has also been observed experimentally in a variety of systems \cite{devoe, bosc1, bosc2,
nellutla, bertaina}. Motivated by such observations, a variant of Markovian quantum master equation has
recently been proposed by Chakrabarti and others \cite{chakrabarti1} which shows that the dissipator has
contribution from the drive as well.  The formulation of the master equation requires explicit introduction
of the fluctuations in the environment which provides a regulator in the dissipator. The presence of the
regulator ensures that the drive-induced dissipation (DID) could be calculated as a simple closed-form
expression. The master equation is named as fluctuation-regulated quantum master equation (FRQME). We note
that in the recent years, FRQME has been used to predict the optimal clock speed of qubit gates and the
nonlinearity of the light shifts \cite{chanda2020, chatterjee2020}.  We use FRQME for the purpose of
including the drive-induced dissipation within the dynamics of the system. More explicitly, in this work, we
propose a dynamical model for quantum measurement that results in the emergence of the Born rule.

If we consider the drive to be much stronger than the system-environment interaction, we would expect that
the system would reach a quasi-steady state with respect to the drive terms, much before the
system-environment coupling influences the system density matrix.  Further, we assume that the drive (in the
form of a pulse) is applied for long enough time for the system to reach a steady state.  As a result,
starting from a pure state, the system ends up with a mixed state and the final density matrix reduces to a
statistical mixture of the eigenstates of the operator part of the drive Hamiltonian with probabilities
being same as that predicted by the Born rule.  The application of the drive acts as a measurement operation
and the operator part of the drive serves as the observable being measured.

\section{Fluctuation-regulated quantum master equation}

In this formulation, one considers the standard settings of a driven open quantum system along with an explicit
introduction of the thermal fluctuation acting on the environment. 
The form of the thermal fluctuations is chosen
to be diagonal in the eigenbasis \{$|\xi_j\rangle$\} of the static Hamiltonian of the
environment, represented by, $\labhl(t) = \sum_j
f_j(t)|\xi_j\rangle\langle\xi_j|$, where $f_j(t)$-s are assumed to be independent, Gaussian,
$\delta$-correlated stochastic variables with zero mean and standard deviation $\kappa$ \cite{chakrabarti1}, \textit{i.e.},
$\overline{f_j(t)} = 0$, 
$\overline{f_j(t_1)f_j(t_2)} = \kappa^2 \delta(t_1-t_2)$. 
This ensures that the fluctuations would destroy the coherences in the environment, but do not change the
equilibrium population distribution of the environment.  Next, we move to the interaction representation
with respect to the static Hamiltonians of the system and the environment, and denote the Hamiltonians with
upright $H$ symbols.  To arrive at the regulator from
the thermal fluctuations, a finite propagator is constructed which is infinitesimal in terms of the drive
and system-environment coupling Hamiltonians (together denoted by $\heff$), but remains finite in the
instances of the fluctuations of the environment.  In order to fulfill this condition, only the first order
contribution of $\heff$ is taken in the construction of the propagator $U(t_1,t)$ within the time interval
$t$ to $t_1$, but we consider many instances of the fluctuation taking place in that interval and retain all
possible higher order terms of $\hl$.  In other words, the timescale of the fluctuations of the environment
is assumed to be much faster compared to the timescale over which the system evolves.  Finally, we get a
finite propagator of the following form,
\begin{equation}
\label{prop2} U(t_1) \approx \ul(t_1) -
i\int_t^{t_1}\heff(t_2)\ul(t_2)dt_2
\end{equation}
where $\ul(t_1) = \mathbb{I} - i\int_t^{t_1}\hl(t_2)\ul(t_2)dt_2$. 

Next the Born approximation \cite{petruccione} is used \textit{i.e.}, at the beginning of the coarse-graining
interval, the total density matrix of the system-environment pair can be factorized into that of the system and the
environment as, $\rho(t) = \rhos(t) \otimes  \rhol$.
This approximation and the assumptions regarding the nature of the
fluctuation provide the desired regulator in the second order under an ensemble average as,
\begin{equation}
\overline{\ul(t_1)\trho(t)\ul^\dagger(t_2)} = \rhos(t) \otimes  \rhol e^{-\frac{1}{2}\kappa^2|t_1-t_2|}.
\end{equation}
A regular coarse-graining procedure \cite{cohen2004}, is subsequently carried out to obtain the fluctuation-regulated 
quantum master equation (FRQME) in the following form:
\begin{widetext}
\begin{eqnarray}
\label{frqme}\frac{d}{dt}{\rhos}(t) = &-& i \; \trl [H_{\rm eff} (t),{\rhos}(t)\otimes \rhol]^{sec} \nonumber \\
&-& \int_0^\infty d\tau\; \trl [H_{\rm eff}(t),[H_{\rm eff} (t-\tau),{\rhos}(t)\otimes \rhol]]^{sec}\; 
e^{-\frac{|\tau|}{\tau_c}}
\end{eqnarray}
\end{widetext}
where, $\tau_c = 2/\kappa^2$ is the characteristic timescale of the decay of the autocorrelation of the fluctuations and  the superscript `\textit{sec}' stands for secular approximation that
involves ignoring the fast oscillating terms in the quantum master equation.  We note that since $\heff$
contains the drive term, hence the DID originates from the double commutator under the integral in the above
equation. 

The FRQME is in Gorini-Kossakowski-Lindblad-Sudarshan (GKLS) form and yields a trace-preserving, completely
positive dynamical map. This FRQME predicts simpler forms of DID, which have been shown to be the absorptive
Kramers-Kronig pairs of the well-known Bloch-Siegert and light shift terms. The predicted nature of DID from
the FRQME has also been verified experimentally \cite{cbEcho2018, chatterjee2020}.

\section{The model}

We consider a strongly-driven system which is weakly coupled to its local environment. We use FRQME to
follow the Markovian dynamics of this system. We note that for our system $\heff$ is given by $\hsl + \hs$.
For a simple Jaynes-Cummings type system-environment coupling, we have $\trl\{\hsl \rho\}=0$ and the FRQME
reduces to,
\begin{eqnarray}
\label{eq_dd}\dot{\rhos} = -i[\hs,\rhos]^{\rm sec} - \ds\rhos - \dsl\rhos 
\end{eqnarray}
where, $\ds$ and $\dsl$ are the Lindbladians from $\hs$ and $\hsl$, respectively, where $\mathcal{D}$
represents the double commutator.  We note that the cross terms between the two
Hamiltonians vanish since $\trl\{\hsl \rho\}=0$. For a strong drive which results in $\ds \gg
\dsl$, it is expected that the system would reach a quasi steady-state with respect to the
commutator and the dissipator $\ds$, and would be influenced by $\dsl$ at a
much later stage (region III), as depicted in the figure 1.

\begin{figure}
\includegraphics[width=6cm]{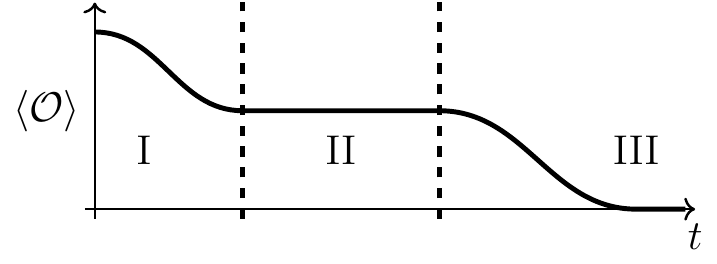}
\caption{A schematic diagram of the nature of the evolution of an observable of the system. In the regions I
and II, the first order commutator and the dissipators $\ds$ dominate the evolution of the system, provided
$\ds \gg \dsl$. In the region III, the dissipators $\ds$ and $\dsl$ together dictate the evolution of the
system.  Region II is a quasi-steady-state where the Born rule emerges.  To analyze the behavior of the
system in the regions I and II, we neglect the dissipator from the system-environment coupling.}
\label{fig-evolve}
\end{figure}

The drive Hamiltonian $\hs$ is chosen to be time-independent and its operator part contains only the
observable of interest.  This choice is made on the ground that if we perform rotating wave approximation
(RWA) on a resonant oscillating drive or use a resonant circularly polarized oscillating drive, it would
result in the same Hamiltonian.

As such, for the strong drive, the equation (\ref{eq_dd}) reduces to the following effective form for the regions I and II,
\begin{eqnarray}
\label{eq5}\frac{d\rhos}{dt} = -i\; [\hs,{\rhos}]\; -\tau_c[\hs,[\hs,{\rhos}]].
\end{eqnarray}
This is the form of FRQME that we shall be using in the remaining part of the manuscript.

Let the eigenvalues and the eigenvectors of $\hs$ be
$\{\lambda_i\}$ and $\{\ket{\phi_i}\}$, respectively. 
Therefore, $\hs$ can be written as, $\hs = \sum_{i} \lambda_i \ket{\phi_i} \bra{\phi_i} $.

Let, $ \ket{\psi_i} = \sum_{j} c_j^i \ket{\phi_j} $ and
\begin{eqnarray*}
\rhos &=&  \sum_i p_i \ket{\psi_i} \bra{\psi_i} \\
&=&  \sum_{j,k} \sum_i p_i c_j^i c_k^{i \; *} \: \ket{\phi_j} \bra{\phi_k} \\
&=&  \sum_{j,k} a_{jk}  \ket{\phi_j} \bra{\phi_k}
\end{eqnarray*}
where $a_{jk}=  \sum_i p_i c_j^i c_k^{i \; *} $.
Rewriting equation (\ref{eq5}) in terms of its $(i,j)^{\rm th}$ elements as,
\begin{equation}\label{rhoij}
\frac{d}{dt} \rhos \vert_{ij} = -i [\hs, \rhos]_{ij} - \tau_c [\hs,[\hs,\rhos]]_{ij}
\end{equation}
Let the $(i,j)^{\rm th}$ element of the density matrix be given by, $(\rhos)_{ij} = \bra{\phi_i} \rhos
\ket{\phi_j} = a_{ij}$.
We can express the equation (\ref{rhoij}) in terms of $a_{ij}$ as,
\begin{eqnarray}\label{a_ij}
\dot{a}_{ij} 
&=& [-i \Delta \lambda_{ij} - \tau_c \Delta \lambda_{ij}^2] a_{ij} 
\end{eqnarray}
where $\Delta \lambda_{ij} = (\lambda_i - \lambda_j)$. 
It is clear that $\dot{a_{ii}} = 0$, that means the diagonal elements do not evolve with time.

The solution of the equation (\ref{a_ij}) is,
\begin{equation}
a_{ij} (t) = a_{ij}(0) e^{-i \Delta \lambda_{ij} t} e^{-\tau_c \Delta \lambda_{ij}^2 t}
\end{equation}

\textbf{Non-degenerate case}: 
If $\lambda_i \neq \lambda_j $, then $a_{ij}({t \rightarrow \infty}) = 0$ and $a_{ii}({t \rightarrow \infty}) = a_{ii}(0) = constant$.
Therefore, all off-diagonal elements will vanish and only diagonal elements will survive in the limit $t \rightarrow \infty$ and the density matrix can be expressed as,
\begin{equation}
\rhos ({t \rightarrow \infty}) = \sum_i \sum_m p_m |c_i^m|^2 \ket{\phi_i}\bra{\phi_i}
\end{equation}

\textbf{Degenerate case}:
If $\lambda_i = \lambda_j $, then $\dot{a_{ij}} = 0 $ and $a_{ij}({t \rightarrow \infty}) = a_{ij}(0) = constant$.
Therefore, both diagonal and off-diagonal elements remain constant in the limit $t \rightarrow \infty$ and the density matrix can be expressed as,

\begin{equation}
\rhos({t \rightarrow \infty})\vert_{i,j} = \sum_m p_m c_i^m c_j^{m \; *}  \ket{\phi_i} \bra{\phi_j}
\end{equation}

The above forms of the density matrix is identical to the form predicted by the Born rule. 

\section{Examples}

We exemplify the emergence of the Born rule for a single qubit system and also for a multi-qubit system. For
the latter, we employ a drive which has a degenerate eigensystem.

\textbf{Single qubit}: For the single qubit system, we begin with the system in a pure state given by,
\begin{equation}
|\psi\rangle =
\cos \frac{\theta}{2} \ket{0}
+ e^{i \varphi}\sin \frac{\theta}{2} \ket{1},
\end{equation}
and the initial density matrix of the system is given by,
$\rho_0 = |\psi\rangle \langle \psi|$.
To emulate the evolution of the system under a strong drive, we apply
a pulse with flip angle $\kappa$ on the system about $y$-axis.
The Hamiltonian corresponding to this pulse is expressed as,
\begin{equation}
\hs = \omega_1 \frac{\sigma_y}{2}
\end{equation}
and the time required to apply this pulse is $\frac{\kappa}{\omega_1}$.

The eigenvectors of the operator part of the drive Hamiltonian $\hs$ are,
\begin{equation}
|\phi_{1}\rangle = \frac{1}{\sqrt{2}} \left(\ket{0} + i\ket{1}\right),\;
|\phi_{2}\rangle = \frac{1}{\sqrt{2}} \left(\ket{0} - i\ket{1}\right),
\end{equation}
and the corresponding non-degenerate eigenvalues are $\pm\omega_1/2$, respectively.
We note that the initial state $\ket{\psi}$ can also be expressed as $\ket{\psi} = \sum_i c_i \ket{\phi_i}$, where, 
$c_1 = \innerp{\phi_1}{\psi} = (\cos\frac{\theta}{2} - ie^{i\varphi}\sin\frac{\theta}{2})/\sqrt{2}$ and
$c_2 = \innerp{\phi_2}{\psi} = (\cos\frac{\theta}{2} + ie^{i\varphi}\sin\frac{\theta}{2})/\sqrt{2}$.

The equation (\ref{eq5})  can be expressed in the Liouville space as follows,
\begin{equation}
\label{eqLiouville}\frac{d\hat{\rhos}}{dt} = 
[-i \hat{\hat{\mathcal{L}}}^{(1)}_{\rm drive} - \tau_c \hat{\hat{\mathcal{L}}}^{(2)}_{\rm drive}]\hat{\rhos} = \hat{\hat{\Gamma}} \hat{\rhos} 
\end{equation}
where, $\hat{\hat{\mathcal{L}}}^{(1)}_{\rm drive}$ is the Liouville superoperator or Liouvillian for the
corresponding $[\hs, \rhos]$ term and $\hat{\hat{\mathcal{L}}}^{(2)}_{\rm drive}$ is the Liouvillian for the
corresponding $[\hs,[\hs,\rhos]]$ term which is responsible for the second order drive-induced dissipation.

Solving this differential equation (\ref{eqLiouville}), we can write the system density matrix at a later time $t$ as, 
\begin{equation}
\label{frqme_red}\hat{\rhos}(t) = e^{\hat{\hat{\Gamma}}t} \hat{\rhos}(0)
\end{equation}
where, $\hat{\rhos}(0)$ is the initial density matrix and $e^{\hat{\hat{\Gamma}}t}$ is the propagator in Liouville space.

We construct the superoperator $\hat{\hat{\Gamma}}$ using equation (\ref{eqLiouville}) and 
construct the Liouville space propagator as, $U =
e^{{\hat{\hat{\Gamma}}} \frac{\kappa}{\omega_1}}$ which acts on the initial state $\rho_0$ to produce the
final density matrix.

When we apply this propagator $U$ on the initial density matrix $\rho_0$, the final density matrix becomes,
\begin{eqnarray}
\rhos = U \rho_0 = 
\begin{pmatrix}
1-e^{-\omega_1 \tau_c  \kappa } a & -i \sin \varphi \sin \theta + e^{-\omega_1 \tau_c  \kappa } b \\
i \sin \varphi \sin \theta + e^{-\omega_1 \tau_c  \kappa } b & 1 + e^{-\omega_1 \tau_c  \kappa } a 
\end{pmatrix},
\end{eqnarray}
where, $a = \left(\sin \theta  \sin \kappa \cos \varphi - \cos \kappa \cos
\theta \right)$, $b = \left( \cos \kappa  \cos \varphi \sin \theta +
\cos \theta \sin \kappa \right)$. The dissipator $\ds$ as described earlier, provides the decaying terms in the above.
In the limit $\omega_1 \tau_c \kappa \rightarrow \infty$, the final density matrix becomes,
\begin{equation}
\rhos = \frac{1}{2} \begin{pmatrix}
1 & -i \sin \theta \sin \varphi \\
i \sin \theta \sin \varphi & 1
\end{pmatrix}
\end{equation}
We can express the above $\rhos$ as,
\begin{equation}
\rhos = \sum_{i=1,2} \vert c_i\vert^2 \ket{\phi_i}\bra{\phi_i}
\end{equation}
where, 
$\vert c_{1} \vert^2 = {\vert\innerp{\phi_{1}}{\psi}}\vert^2 = \frac{1}{2}(1 + \sin \theta \sin \varphi)$, and
$\vert c_{2} \vert^2 = {\vert\innerp{\phi_{2}}{\psi}}\vert^2 = \frac{1}{2}(1 - \sin \theta \sin \varphi)$.
This defines a mixed state density matrix of the eigenstates of the drive.
As such, the drive projects the initial state of the system onto one of its eigenstates with probabilities given by $|c_1|^2$ and $|c_2|^2$, respectively.
Therefore, our result agrees with the Born rule.
Figure 2 shows a schematic diagram of the evolution of the system on a Bloch sphere, with the classical
paths described by colored arrows. The system moves from a pure state to a mixed state.

\begin{figure}[htb]
\includegraphics[width=6cm]{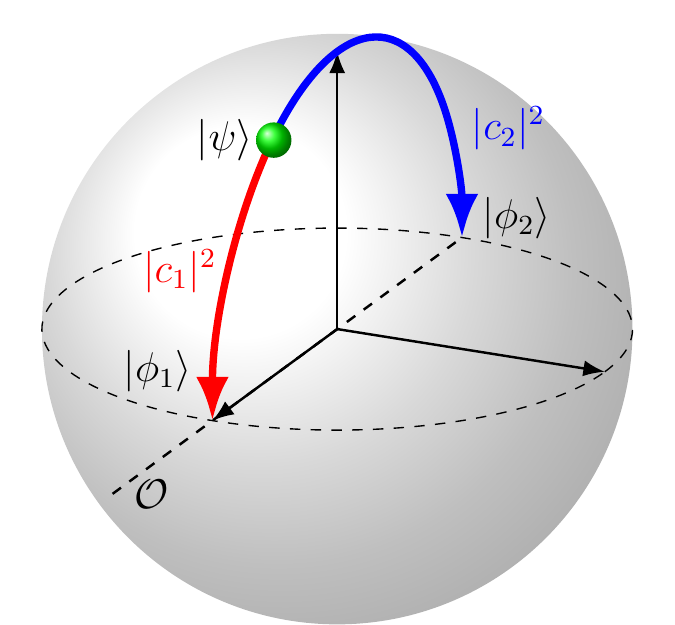}
\caption{Schematic depiction of the journey to the final states starting from $\ket{\psi} = \sum_i c_i
\ket{\phi_i}$ (the green sphere) where $\{\phi_i\}$ are the eigenstates of the observable $\mathcal{O}$.
$\vert c_i\vert^2$ is the probabilities of the path $\ket{\psi}\rightarrow\ket{\phi_i}$. The final
state is a mixed state density matrix in accordance with the Born rule.}
\end{figure}

\textbf{Degenerate observable and multi-qubit}: 
We extend our analysis to an observable with degenerate eigensystem. We choose a two-qubit system in an
entangled state given by,
\begin{equation}
\ket{\psi} =
\frac{1}{\sqrt{2}}(\ket{00}+\ket{11})
\end{equation}
We intend to measure $\sigma_y\otimes \mathbb{I}$ on this system, such that the measurement takes place only
on the first qubit. The eigenvectors of this observable are,
\begin{eqnarray}
\ket{\phi_1} &=& \frac{1}{\sqrt{2}}(\ket{01}-i\ket{11}), \;
\ket{\phi_2} = \frac{1}{\sqrt{2}}(\ket{00}-i\ket{10}) \\
\ket{\phi_3} &=& \frac{1}{\sqrt{2}}(\ket{01}+i\ket{11}), \;
\ket{\phi_4} = \frac{1}{\sqrt{2}}(\ket{00}+i\ket{10}) \\
\end{eqnarray}
and the corresponding eigenvalues are, $-1, -1, 1 ,1$, respectively. 

The initial density matrix of the system is given by,
$\rho_0 = \ket{\psi}\bra{\psi}$. 
Like the single qubit case, the operation of the observable is emulated by
a pulse with flip angle $\kappa$ on the first qubit about $y$-axis .
The Hamiltonian corresponding to this pulse is expressed as,
\begin{equation}
\hs = \omega_1 \frac{\sigma_y}{2} \otimes \mathbb{I}
\end{equation}
and the time required to apply this pulse is $\kappa/\omega_1$.

The system evolves under the Liouvillian obtained using the equation (\ref{eqLiouville}) and the final density matrix assumes
the form,
\begin{equation}
\rhos = U \rho_0 = \frac{1}{4}
\begin{pmatrix} (1 + e^{-2 \omega_1 \tau_c  \kappa } C) & -e^{-2 \omega_1 \tau_c  \kappa } S & e^{-2 \omega_1
\tau_c  \kappa } S &
(1+ e^{-2 \omega_1 \tau_c  \kappa } C) \\
-e^{-2 \omega_1 \tau_c  \kappa } S & (1 - e^{-2 \omega_1 \tau_c  \kappa } C) & (-1+ e^{-2 \omega_1 \tau_c
\kappa } C) & -e^{-2 \omega_1 \tau_c
\kappa } S \\
e^{-2 \omega_1 \tau_c  \kappa } S & (-1 + e^{-2 \omega_1 \tau_c  \kappa } C) & (1 - e^{-2 \omega_1 \tau_c
\kappa } C) & e^{-2 \omega_1 \tau_c
\kappa } S \\
(1+ e^{-2 \omega_1 \tau_c  \kappa } C) & -e^{-2 \omega_1 \tau_c  \kappa } S & e^{-2 \omega_1 \tau_c  \kappa } S &
(1+ e^{-2 \omega_1 \tau_c  \kappa } C)
\end{pmatrix}.
\end{equation}
where, $S = \sin \kappa$, and $C = \cos \kappa$.  

In the limit $\omega_1 \tau_c \kappa \rightarrow \infty$, the final density matrix becomes,
\begin{equation}
\rhos = \frac{1}{4}
\begin{pmatrix}
1 & 0 & 0 & 1 \\
0 & 1 & -1 & 0 \\
0 & -1 & 1 & 0 \\
1 & 0 & 0 & 1
\end{pmatrix}.
\end{equation}
We can express the above $\rhos$ as,
\begin{equation}
\rhos = \sum_{i,j \in \{1,2\}} c_i c_j^*  \ket{\phi_i}\bra{\phi_j} + \sum_{i,j \in \{3,4\}} c_i c_j^*  \ket{\phi_i}\bra{\phi_j}
\end{equation}
where,
$c_1= {\innerp{\phi_1}{\psi}} = 
\frac{i}{2}$, $c_2= {\innerp{\phi_2}{\psi}}=\frac{1}{2}$, $c_3= {\innerp{\phi_3}{\psi}}= -\frac{i}{2}$,
$c_4= {\innerp{\phi_4}{\psi}}=\frac{1}{2}$.

The off-diagonal terms will be present in the final form of $\rhos$ because of the degeneracy between
$\ket{\phi_1}$, $\ket{\phi_2}$ and $\ket{\phi_3}$, $\ket{\phi_4}$.
We can rewrite this as a mixture of the linear superpositions of the states in the following form:
\begin{eqnarray*}
\rhos = (|c_1|^2 + |c_2|^2 ) \ket{\Phi_1}\bra{\Phi_1} + (|c_3|^2 + |c_4|^2 ) \ket{\Phi_2}\bra{\Phi_2} 
\end{eqnarray*}
where $\ket{\Phi_1} = (c_1 \ket{\phi_1} + c_2 \ket{\phi_2})/\sqrt{|c_1|^2 + |c_2|^2}$ and $\ket{\Phi_2} =
(c_3 \ket{\phi_3} + c_4 \ket{\phi_4})/\sqrt{|c_3|^2 + |c_4|^2}$ are the normalized superposition states.
The application of the drive causes the system to collapse in the degenerate eigensubspaces formed by
$\ket{\phi_1}$, $\ket{\phi_2}$ and $\ket{\phi_3}$, $\ket{\phi_4}$ with probabilities being $(|c_1|^2 + |c_2|^2 )$ and $(|c_3|^2 + |c_4|^2 )$, respectively.
Therefore, it has been verified that the results of the present example is also consistent with the Born rule.

\section{Discussions}

\begin{figure}[htb]
\includegraphics[width=5cm]{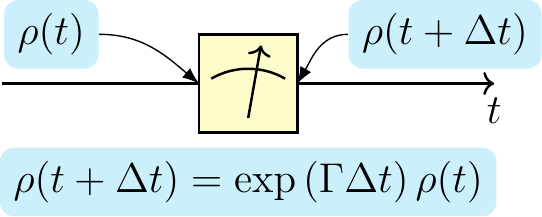}
\caption{A schematic diagram of the model of the measurement. After the measurement, $\rho(t + \Delta t)$ is
of the form predicted by the Born rule. $\Gamma \Delta t$ should be sufficiently large for the system to
reach a steady state.}
\label{fig-rho} 
\end{figure}

The principal feature of our model is that the first-order effect of the drive in tandem with the
drive-induced dissipation lead the system to a mixed state. It is assumed that the drive is stronger than
the system-environment coupling and as such the dissipator from the drive acts before the regular dissipator
from the system-environment coupling acts. Also, we require that the drive is applied for sufficiently long
duration to reach a quasi-steady-state which is identical to the state predicted by the Born rule. We note
that the focus is on the creation of the mixed state through an irreversible dynamics. The lack of an
explicit apparatus means that we may not be able to \emph{register} the outcome of a specific measurement,
but that is not what this model intends to achieve. Even if one does not register the outcome of a
measurement, a probabilistic mixed state description is reached and one can apply this repeatedly to model
the measurement many times without having to \emph{reset} the apparatus, as shown in the schematics in
figure \ref{fig-rho}. This is one of the major advantages of this model.

One simplifying assumption in this model was $\ds \gg \dsl$. In a situation, where the assumption fails, one
would expect a competition between the decoherence due to the drive and system-environment interaction. We
have shown that similar effect gives rise to the existence of the optimal clock speed for qubit gates in open
quantum systems \cite{chanda2020}.  As a result, the system would end up in a different mixed state which
might not be decomposed into the eigenstates of the drive since the system would not be diagonal in the
eigen representation of the drive anymore. So, the Born rule might not be obtained in that case.  We
understand that the Born rule is an asymptotic solution. The system evolves to a steady state no matter
whether the drive is strong. If the drive is stronger than system-environment coupling, then only it takes
the system to the eigenstates as speculated by the measurement postulate, else the system reaches a
different steady state not prompted by the Born rule.  We are not actually ignoring the system-environment
coupling; rather we confine the FRQME to the drive-drive term only so as to study the nonunitarity induced
in the system solely caused by the drive.  The DID is scaled by the term  $\tau_c$ that carries a signature
from the environmental fluctuations and determines how fast one can reach a steady state. So, higher
$\tau_c$ means faster collapse. The emulation of the Born rule is favored at lower temperature.

In the year 1991, Milburn proposed a model for intrinsic decoherence based on a simple modification of a
unitary Schr\"odinger evolution and derived an equation of motion for the density matrix of closed quantum
systems as a substitution for Schr\"odinger equation \cite{milburn}. This is known as Milburn equation.  For
sufficiently small fundamental time steps with terms up to second order being considered, Milburn equation
reduces to FRQME of the form given by equation (\ref{eq5}).  Few years later, Bu$\check{\rm z}$ek and
Kon\^opka applied Milburn equation to an open quantum system consisting of a two-level atom interacting with
single- and multi- mode electromagnetic field \cite{buzek}.  They reported that for very strong
system-environment coupling, Rabi oscillation is completely suppressed and the system collapses to a
statistical mixture of the ground and excited states.  We note that their study focuses on the overdamped
nature of the system, but is not a model of measurement process.  On the other hand, our model emulates the
collapse part of the measurement by presenting a dynamical treatment of how an open quantum system behaves
when a drive is applied on it. As per our model, measuring an observable is equivalent to evolving the
system under that observable (which happens to be the drive) with a large amplitude, \emph{i.e.} evolving
the system strongly under the observable. This naturally leaves the system in a mixed state formed by the
eigenstates of the observable with probabilities given by the Born rule.

\section{Conclusion}

In this work, we demonstrate that the drive-induced dissipation from a strong drive can result in
the emergence of the Born rule in a system weakly coupled to the environment. We assume that the
dissipator from the drive is much larger than the dissipator from the system-environment coupling. The
resulting dynamics is best analyzed in the eigenbasis of the drive, where, the evolution destroys the
coherences. Thus the final density matrix is in a mixed state and is diagonal in this
representation for a non-degenerate observable. For an observable with the degenerate eigenvalues, the
coherences in the degenerate subspace survives in conformity with the Born rule. This dynamic model emulates
the Born rule and could be used repeatedly on a system. 

In the quantum information processing, often measurements are included in the quantum circuits. Such
measurements could be on multiple qubits and occur more than once in the circuit. Our model could be very
useful in simulating the dynamical behavior of a realistic open quantum system which has multiple occurrence
of measurements. One would obtain the mixed state representation of the system at the end of the circuit
operation, with the added advantage of not having to \emph{reset} the apparatus. As examples, we have
demonstrated the measurement operation for a single- and multi-qubit arrangements.

\end{document}